\documentclass[12pt]{article}
\usepackage[varg]{txfonts}

\usepackage{epsf,rotating,latexsym,amssymb} 
\usepackage{epsfig}   
\usepackage{epstopdf} 
\usepackage{rotating,latexsym,amssymb}           
\usepackage{mathptmx} 
\usepackage{graphicx}
\usepackage{times}
\usepackage{natbib}
\usepackage{lineno}       
\usepackage{color}
\newcommand{\kms}{$\mathrm {km\,s^{-1}}$}

\begin{document}
\bibliographystyle{apalike}

\hfill {\em Accepted for publication in Astronomy \& Astrophysics, 2019}
\medskip

\noindent
\Large{\bf Interstellar Dust in the  Solar System: Model versus  In-Situ Spacecraft Data} 

\normalsize

\medskip
\medskip

\noindent
Harald Kr\"uger (krueger@mps.mpg.de)$^{1}$;  
Peter Strub$^{1,2}$;  
Nicolas Altobelli$^{3}$; 
Veerle Sterken$^{4}$; \\ 
Ralf Srama$^{2}$; 
Eberhard Gr\"un$^{5}$ 

\bigskip


\noindent
1) Max-Planck-Institut f\"ur Sonnensystemforschung, G\"ottingen, Germany;  
2) Institut f\"ur Raumfahrtsysteme, Universit\"at Stuttgart, Germany; 
3) European Space Agency, ESAC, Madrid, Spain; 
4) Institute of Applied Physics, University of Bern, Switzerland; 
5) Max-Planck-Institut f\"ur Kernphysik, Heidelberg, Germany

\abstract

In the early 1990s, contemporary interstellar dust penetrating deep into the heliosphere was identified 
with the in-situ dust detector on board the Ulysses spacecraft. Later on, interstellar dust was also 
identified in the
data sets measured with  dust instruments on board Galileo, Cassini and Helios. 
 Ulysses monitored the interstellar dust stream at high ecliptic latitudes 
for about 16 years.  
The three other spacecraft data sets were obtained 
in the ecliptic plane and cover much shorter time intervals.
We compare  in-situ interstellar 
dust measurements obtained  with these four spacecrafts, published in the literature, 
with  predictions of a state-of-the-art
model for the dynamics of interstellar dust in the inner solar system (Interplanetary Meteoroid 
environment for EXploration, IMEX), 
in order to test the reliability of the model predictions.  
Micrometer and sub-micrometer sized dust particles are subject to solar gravity and radiation pressure 
as well as to the Lorentz force on a charged dust particle moving through the Interplanetary Magnetic Field, 
leading to a complex size dependent 
flow pattern of interstellar dust in the planetary system. The  IMEX model was 
calibrated with the Ulysses interstellar dust measurements and includes these relevant 
forces. We study the time-resolved flux and mass distribution 
of interstellar dust in the  solar system. 
The IMEX model agrees with the spacecraft measurements within a factor of 2 to 3, also for time intervals 
and spatial regions not covered by the original model calibration with the Ulysses data set. It  
usually underestimates the dust fluxes measured by the space missions which were not used for the model calibration,
i.e. Galileo, Cassini and Helios. 
IMEX is a unique time-dependent model for the prediction of interstellar dust fluxes and
mass distributions for the inner and outer solar system. The model is suited to study 
dust detection conditions for past and future space missions.

\section{Introduction}

Interstellar dust became a topic of astrophysical research
in the early 1930s when astronomers realized the extinction
of starlight in the interstellar medium (ISM). At that time, information about 
dust in the ISM could only be obtained  by astronomical observations.
With the advent of dust detectors onboard spacecraft in the 1970s, it became possible to 
investigate dust particles in-situ, and the analysis of data
obtained with the dust instruments flown on a couple of spacecraft suggested 
that interstellar dust  can cross the heliospheric boundary and penetrate deep
into the heliosphere \citep[][see \citet{krueger2009a} for a review]{bertaux1976,wolf1976}. 
Later on, in the 1990s, this 
was undoubtedly demonstrated  by the Ulysses
spacecraft: the Ulysses 
dust detector, which measured mass, speed and approach direction of the 
impacting particles, identified interstellar particles with radius
above $\mathrm{0.1\,\mu m}$ sweeping through the heliosphere 
\citep{gruen1993a,gruen1994a,gruen1995a}. These particles originated from the local interstellar cloud 
(LIC) surrounding our solar system \citep{frisch1999a}, thus
providing an opportunity to probe dust from the LIC.

The motion of the heliosphere with respect to this cloud causes an inflow of 
interstellar dust into the heliosphere from a direction of $259^{\circ}$ ecliptic longitude and 
$8^{\circ}$ latitude \citep{landgraf1998a,frisch1999a,strub2015} with an inflow speed 
of $\mathrm{26\,km\,s^{-1}}$ \citep{gruen1994a,krueger2015a}. Within the measurement accuracy, 
the average dust inflow direction is co-aligned with the interstellar neutral helium flow 
\citep{witte1996,witte2004b,wood2015}\footnote{Working values of a speed of $\mathrm{25.4 km\,s^{-1}}$ with directions 
from 255.7$^\circ$ ecliptic longitude and +5.1$^\circ$ ecliptic latitude were suggested from 
Energetic Neutral Atom measurements by the Interstellar Boundary EXplorer mission 
(IBEX) by~\citet{mccomas2015b} and recently confirmed by \citet{swaczyna2018}.}. 
The interstellar dust flow persists at high ecliptic latitudes 
above and below the ecliptic plane and 
even over the poles of the Sun, whereas interplanetary dust is strongly depleted at high 
latitudes \citep{gruen1997a}. 

The Ulysses interstellar dust measurements
were  confirmed by the Galileo \citep{baguhl1996,altobelli2005a} and Cassini spacecraft 
\citep{altobelli2003,altobelli2007,altobelli2016}, 
and interstellar impactors were also identified in the Helios dust data \citep{altobelli2006}. 
In 2006, the Stardust mission  
successfully brought a sample of collected interstellar particles to Earth \citep{westphal2014b}.
Finally, measurements by the radio and plasma wave instruments on board the STEREO and WIND 
spacecraft were explained by
interstellar dust  \citep{belheouane2012,malaspina2016}, although this interpretation
was recently called into question \citep{kellogg2018}.

Measurements of interstellar dust inside the planetary system now provide a 
window for the study of solid interstellar matter at our doorstep \citep{frisch1999a}. However, 
the flow of the interstellar particles in the heliosphere is governed by two fundamental effects: 
(1) the combined gravitational force and the radiation pressure force of the Sun, and (2) the Lorentz force 
acting on
 a charged particle moving through the solar magnetic field ''frozen`` into the solar wind. The former effect 
can be described as a multiplication of the gravitational force by a constant factor $(1 - \beta$), where the radiation pressure factor $\beta = |{\bf F}_{rad}|/|{\bf F}_{grav}|$ is a function of particle composition, size and morphology. Interstellar particles approach the Sun on hyperbolic trajectories, leading to either a radially symmetric focussing ($\beta  < 1$) or defocussing ($\beta > 1$) downstream of the Sun which is constant in time \citep{bertaux1976,landgraf2000b,sterken2012a}. Particle sizes observed by the Ulysses dust detector typically range from approximately $\mathrm{0.1\,\mu m}$ to several micrometers, corresponding to $\mathrm{0 \lesssim \beta \lesssim 1.9}$~\citep{kimura2003b,landgraf1999a}\footnote{\citet{landgraf1999a} found a range of  $1.4 < \beta < 1.8$  from Ulysses measurements, and~\citet{kimura2003b} found values for $\beta$ between 0 and 1.9}.
A detailed description of the forces acting on the particles and the resulting general interstellar dust flow characteristics was given by \citet{sterken2012a}. Reviews about interstellar dust measurements 
in the solar system were recently given by \citet{mann2010} and \citet{sterken2019}.

The interplanetary magnetic field (IMF) shows systematic variations with time, including the 25-day 
solar rotation 
and the 22-year solar magnetic cycle, as well as local deviations caused by disturbances in the 
interplanetary magnetic field, due to, e.g. coronal mass ejections  (CMEs). The dust particles in interplanetary 
space are typically 
charged to an 
equilibrium potential of +5~V \citep{kempf2004}.  Small particles have a 
higher charge-to-mass ratio, hence their dynamics is more sensitive to the interplanetary magnetic 
field. The major effect of the magnetic field on the charged interstellar dust  is a 
focussing and defocussing relative to the solar equatorial plane  with the 22-year 
magnetic cycle of the Sun \citep{landgraf2000b,landgraf2003,sterken2012a,sterken2013a}. 
Modifications of the particle dynamics by solar radiation pressure and the Lorentz force acting on 
 charged dust particles
have to be taken into account for a proper interpolation of the interstellar dust properties
to the interstellar medium outside the heliosphere where these particles originate from.

Results of interstellar dust measurements and simulations (including mass distributions) 
from Galileo and Ulysses were compared and studied by \citet{landgraf2000a}. A first comparison 
of the interstellar dust data obtained with four spacecraft, i.e. Ulysses, Galileo, Cassini and Helios,
was performed by \citet{altobelli2005b}. The results showed a decrease of the
measured flux in the inner solar system which was attributed to heliospheric filtering. 
However, no  comparison with a detailed dynamical model 
for all four missions was possible at the time.
Only for the Galileo and Ulysses (partial) data sets~\citep{landgraf1998a} and the (complete) Ulysses data set~\citep{sterken2015}, detailed comparisons have been made. Here we use these same data and 
compare them with simulation results obtained from our latest model for the dynamics of 
interstellar dust in the solar system \citep{landgraf2000b,sterken2013a,strub2019}. 

In  Section~\ref{sec:data} we briefly present the interstellar dust measurements obtained 
by the Helios, Cassini, Galileo and Ulysses missions. For
a comprehensive description of the data analysis, in particular the identification scheme 
for the interstellar impactors, the reader is referred to the publications by 
\citet{altobelli2003,altobelli2005a,altobelli2006,altobelli2016} for Helios, Galileo and Cassini, 
and \citet{strub2015} for the Ulysses data. In Section~\ref{sec:model} 
we present our modelling results and compare them with the in-situ measurements.
Section~\ref{sec:discussion} is a Discussion and in Section~\ref{sec:conclusions} we
summarise our conclusions.

\section{In-Situ Spacecraft Dust Data}

\label{sec:data}

The physical mechanism most generally utilized in modern spaceborne 
in-situ dust detectors is based on the measurement
of the electric charge generated upon impact of a fast 
projectile on to a solid target (impact ionization).    
It yields the highest 
sensitivity for the detection of dust particles in space
\citep{fechtig1978,auer2001}. The impact can be detected by several independent measurements on different 
instrument channels (multi-coincidence detection) which allows for a reliable dust
impact detection and identification of noise events \citep{gruen1992a}. 
The electrical charge generated upon  impact 
can be empirically calibrated to provide the impact speed and the mass of the 
particle \citep{goeller1985}. In combination with a time-of-flight 
mass spectrometer, an impact ionisation detector can measure the chemical composition of the 
impacting particle \citep{srama2004}. 

In this work we use dust data obtained by impact
ionization dust detectors on board the spacecraft  Helios, Galileo, Cassini and Ulysses 
\citep{dietzel1973,gruen1981a,gruen1992a,gruen1992b,srama2004}. Impacts of interstellar
dust particles in these data sets were identified by \citet{altobelli2006,altobelli2005a,altobelli2003,altobelli2016}
and \citet{strub2015}. We do not consider dust measurements with other detection
techniques here because we want to keep the data set as consistent as possible. Different detection techniques are
usually connected with individual systematic uncertainties, e.g. due to 
mass calibration or instrument detection threshold, increasing the overall uncertainty. 

When a dust particle strikes a solid target with high speed ($\gg \mathrm{1\,km\,s^{-1}}$),
it forms a crater in the target and releases ejecta composed of both particle
and target material. The ejecta consist of positive and negative ions, 
electrons, and neutral atoms and molecules originating from 
both projectile and target. Because of its high internal pressure 
(up to 5 TPa), the ejecta cloud expands rapidly into the surrounding vacuum. 

The sensors consist of a metal  plate target
and a collector (e.g. a metal grid) for either the ions or 
electrons of the impact plasma. 
Different
electric potentials applied to the target plate and the
collector generate an electric field, separating the
positively and negatively charged ions. 
Charge-sensitive amplifiers coupled to both the target
plate and the collector register independently, but 
simultaneously, an impacting dust particle. The total
amount of charge, $Q_{imp}$, collected on each channel
is a function of mass $m_d$ and impact speed $v_d$ of the 
particle as well as the particle's composition. $Q_{imp}$ 
can be described by the empirical law 
\begin{equation}
   Q_{imp} = K \, m_d^{\alpha}\, v_d^{\gamma} ,         \label{eq_charge}
\end{equation}
with $\alpha \simeq 1$ and $ 1.5 \lesssim \gamma  \lesssim 5.5$ in
the speed range 2 \kms\ $\lesssim v \lesssim $ 70 \kms\ 
\citep{auer2001,stuebig2002}. $K$ depends on the sensor geometry and the  
signal processing by the instrument 
electronics. In particular, for constant impact speed, the charge 
generated upon impact
is proportional to the particle mass \citep{goeller1985}.

\begin{table*}[h]
{\tiny
      \caption{ Characteristics of the individual spacecraft measurements.}
        \label{tab:data}
      \begin{tabular}{lcccccccccc} 
 \hline
 \hline
Mission/       &  Start Time   &     End Time &   Range   &Impact& $^{\dagger}Q_{imp}/m_d$ & \multicolumn{3}{c}{$^{\dagger \dagger}$ Detection Threshold} &  $^{\ast}$N  & $^{\ast}$Average flux \\
Interval        &              &              &           & Speed&                   &      Charge        &       Mass         &    Radius            &                 &      \\
                &  [year-doy]  &    [year-doy]&   [AU]    & $v_d$ [\kms]&      [C/kg]       &  $Q_{imp}$   [C]       &  $m_d$   [kg]      & $r_d$ [$\mathrm{\mu m}$] & & [$\mathrm{m^{-2}\,s^{-1}}$]  \\
    (1)         &    (2)       &      (3)     &     (4)   &  (5) &      (6)          &   (7)              &     (8)            &            (9)           &    (10)   &   (11)  \\
\hline
Helios   & & & & & & & & & &\\
HEL$^{\ast \ast}$ &  1974-353  &  1980-002    & 0.3 -- 1.0& 60   & $2.6\cdot 10^{3}$ & $2\cdot 10^{-12}$  & $7 \cdot 10^{-16}$&   0.37      &     27 &$(2.6\pm 0.3) \cdot 10^{-6}$  \\
         & & & & & & & & & & \\[-1.5ex]
Galileo  & & & & & & & & & &\\
GLL1                  &  1990-001 &  1990-190 & 0.7 -- 1.2& 50   & $5.0\cdot 10^{4}$ & $2 \cdot 10^{-12}$ & $4\cdot 10^{-17}$ &    0.14     &  21   & $(7.0\pm 1.5) \cdot 10^{-5}$ \\
GLL2                  &  1991-056 &  1991-123 & 1.0 -- 1.4& 50   & $5.0\cdot 10^{4}$ & $2 \cdot 10^{-12}$ & $4\cdot 10^{-17}$ &    0.14     &  13   &   \\
GLL3                  &  1991-228 &  1991-340 & 1.9 -- 2.2& 30   & $6.5\cdot 10^{3}$ & $1 \cdot 10^{-13}$ & $2\cdot 10^{-17}$ &    0.11     &  19   & \raisebox{1.3ex}[-1.3ex]{$(9.5\pm 1.5) \cdot 10^{-5}$}  \\
GLL4                  &  1993-005 &  1993-181 & 1.2 -- 2.5& 50   & $5.0\cdot 10^{4}$ & $2 \cdot 10^{-12}$ & $4\cdot 10^{-17}$ &    0.14     &  22   & $(3.5\pm 0.8) \cdot 10^{-5}$   \\
GLL5                  &  1993-182 &  1993-365 & 2.5 -- 3.5& 35   & $1.1\cdot 10^{4}$ & $1 \cdot 10^{-13}$ & $9\cdot 10^{-18}$ &    0.09     &  41   & $(8.0\pm 1.0) \cdot 10^{-5}$    \\
         & & & & & & & & & &\\[-1.5ex]
Cassini  & & & & & & & & & &\\
CAS1                   &  1999-081 &  1999-181 & 0.7 -- 1.2& 45   & $6.0\cdot 10^{4}$ & $3 \cdot 10^{-12}$ & $5\cdot 10^{-17}$ &    0.15     &  14   & $(2.5 \pm 0.5)\cdot 10^{-5}$ \\
CAS2$^{\ast\ast\ast}$ &  2004-183 &  2013-364 & 9.1 -- 9.9 & 30 & $8.0\cdot 10^{3}$ & $1 \cdot 10^{-15}$ & $5\cdot 10^{-18}$ & $0.07$&  36 &  $(1.5 \pm 0.5)\cdot 10^{-4}$ \\ 
         & & & & & & & & & & \\[-1.5ex]
Ulysses  & & & & & & & & & & \\
ULS1                  &  1992-245 &  1994-131 & 3.0 -- 5.0& 30   & $6.5\cdot 10^{3}$ & $1 \cdot 10^{-13}$ & $2\cdot 10^{-17}$ &   0.11     &  116   & $(7.7\pm 2.0) \cdot 10^{-5}$ \\
ULS2                  &  1995-166 &  1996-131 & 1.9 -- 3.7& 30   & $6.5\cdot 10^{3}$ & $1 \cdot 10^{-13}$ & $2\cdot 10^{-17}$ &   0.11     &   39   &$(5.3\pm 1.7) \cdot 10^{-5}$ \\
ULS3                  &  1996-131 &  2000-131 & 3.7 -- 5.4& 30   & $6.5\cdot 10^{3}$ & $1 \cdot 10^{-13}$ & $2\cdot 10^{-17}$ &   0.11     &   94   & $(2.9\pm 1.1) \cdot 10^{-5}$  \\
ULS4                  &  2002-131 &  2002-363 & 3.5 -- 4.4& 25   & $4.0\cdot 10^{3}$ & $1 \cdot 10^{-13}$ & $3\cdot 10^{-17}$ &   0.13     &   37   & $(1.1\pm0.3) \cdot 10^{-4}$\\
ULS5                  &  2005-245 &  2006-245 & 3.2 -- 4.9& 30   & $6.5\cdot 10^{3}$ & $1 \cdot 10^{-13}$ & $2\cdot 10^{-17}$ &   0.11     &   79   & $(1.1\pm 0.2) \cdot 10^{-4}$ \\
         \hline
                  & & & & & & & & & & \\[-1.5ex]
          \end{tabular}
          }

{\small          
{\bf Notes.} Spacecraft (column~1), measurement periods (columns~2 and 3), heliocentric distance range (column~4), 
      average interstellar dust impact speed derived from the model (column~5), charge-to-mass ratio  from 
      instrument calibration (column~6),  detection thresholds (columns~7 to 9), number of identified interstellar particles (column~10), 
      and average interstellar dust fluxes (column~11).

$^{\ast}$ \citet{altobelli2006}, \citet{altobelli2005a}, \citet{altobelli2003}, \citet{strub2015}  

$^{\ast \ast}$ For Helios we have considered only impacts when the true anomaly angle of the spacecraft was in the range  $-180^{\circ} < \nu < 90^{\circ}$ consistent with \citet{altobelli2006}.

$^{\ast\ast\ast}$ Cassini CDA was not pointing into the direction of interstellar dust continuously during this time interval \citep{altobelli2016}

$^{\dagger}$ \citet{gruen1981a,gruen1995a,stuebig2002}
   
   $^{\dagger \dagger}$  Detection threshold based on the identification scheme for interstellar particles. 
Throughout this paper we calculate particle radii from the measured masses by assuming a spherical particle 
shape and a density typical of astronomical silicates $\mathrm{\rho_d  = 3300\,kg\,m^{-3}}$ 
\citep{kimura1999}. The particle radius is  given by
\begin{equation}
r_d = \sqrt[3]{\frac{3m_d}{4\pi \rho_d}}, \label{equ_2} 
\end{equation}
where $m_d$ is the dust particle mass derived from the instrument calibration.  
}

\end{table*}

The instrument sensitivity, expressed by the parameter $K$, is determined 
by the technical detection threshold for the impact charge measurement, which is 
about $\mathrm{10^{-14}\,C}$ for Ulysses, Galileo and Helios, and $\mathrm{10^{-15}\,C}$ for Cassini,
respectively. However, the interstellar impactors had to 
 be separated from interplanetary particle impacts which was usually done by their impact direction
 and impact speed (or generated impact charge). This led to specific identification criteria
 for the interstellar particles in the data sets of the four space missions, and thus  in most cases to 
 less sensitive  
  detection thresholds than the technically 
 detectable 
 threshold.  The measurement periods of the different space missions, details of the particle 
 identification schemes and 
 derived dust fluxes are summarized in Table~\ref{tab:data}.

The particle speed can be determined from the rise times of the individual
charge signals \citep{goeller1989}. For a given impact speed the signal strength also moderately depends
on the particle material and  on the impact angle. Neither the
particle material nor the impact angle are known for an impinging
particle. Therefore, averaged calibration curves have to be used to 
obtain impact speeds, assuming that the materials used for 
calibration represent cosmic dust particles \citep{gruen1995a}. 
The typical accuracy of the derived speed $v_d$ is a factor of 2.

Once the particle speed has been determined, its charge-to-mass ratio $Q_{imp}/m_d$ generated upon impact 
can be derived from empirical calibration curves. These were obtained  
from impact experiments at the dust 
accelerator facility at Max-Planck-Institut f\"ur Kernphysik, Heidelberg 
\citep[$\gamma$ in Equation~\ref{eq_charge};][]{gruen1981a,gruen1995a,stuebig2002,srama2004,srama2009}. In the next step, 
the particle mass can be derived from the calibrated impact charge-to-mass ratio and the
measured impact charges. 
If the speed is
well determined, the mass can also be derived with a
higher accuracy. The typical uncertainty in the  derived mass
$m_d$ is a factor of 10. 

 Given that the
charge-to-mass ratio $Q_{imp}/m_d$  strongly depends on particle impact speed, we had to assume a
 speed in order to derive the particle mass from the measured impact charge  in the spacecraft data
(Equation~\ref{eq_charge}). For our analysis we took  average particle speeds from the model
for the measurement time interval considered (Section~\ref{sec:model}), 
and the corresponding $Q_{imp}/m_d$  listed in Table~\ref{tab:data}.

\begin{figure}[tb]
		\includegraphics[width=0.68\textwidth]{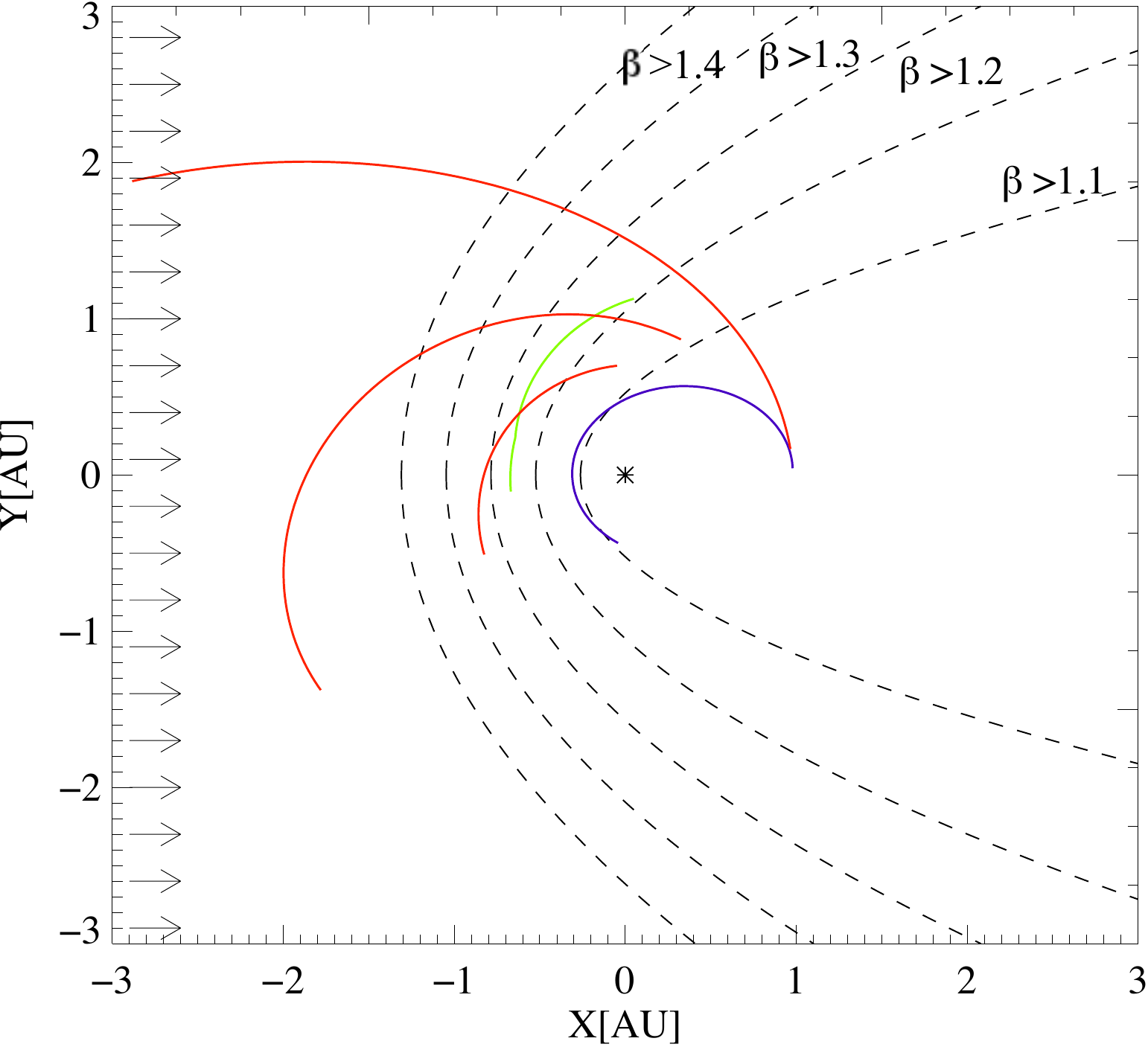}
	\caption{Trajectory segments of Helios (blue), Galileo (red), and Cassini (green) which were favorable for the 
	identification of interstellar dust. The Sun is at the center and the 
	X-Y plane is the ecliptic plane. Vernal equinox is 
	approximately towards the -Y direction, and the nominal interstellar dust flow direction is indicated 
	by arrows. Different $\beta$ avoidance cones are indicated by dashed lines. See text for details.
	  }
	\label{fig:traj_1}
\end{figure}

\subsection{Helios}

The Helios~1 spacecraft (we refer only to Helios~1 throughout this paper) 
 was launched into a 
heliocentric orbit on 10 December 1974,  with perihelion and aphelion distances of 0.3~AU and 1.0~AU, 
respectively (Figure~\ref{fig:traj_1}). 
The spacecraft was spin-stabilized with
a spin axis pointing normal to the ecliptic plane and a spin period of one second. It carried two dust instruments, the
ecliptic sensor which was exposed to sunlight, and the south sensor which was shielded  by the
spacecraft from direct sunlight \citep{dietzel1973,fechtig1978,gruen1981a,altobelli2006}. 

Between 19 December 1974 and 2 January 1980 the Helios sensors transmitted the data of 235 dust impacts 
to Earth \citep{gruen1981a}. Interstellar impactors could only be separated from interplanetary dust particles if their
impact charge exceeded 
$\mathrm{2\cdot 10^{-12}\, C}$ and only during limited periods of the Helios orbit. 
This led to the identification of 27  
interstellar impactors during ten orbits of Helios around the Sun when the spacecraft's true anomaly angle 
$\nu$ was in the range $-180^{\circ} < \nu < 90^{\circ}$ 
\citep{altobelli2006}. Calibration parameters $K=4.07 \cdot 10^{-5}$ (mass taken in gram) and $\gamma =2.7$ \citep{gruen1981a} were used, and the
derived particle masses were mostly
in the range $\mathrm{10^{-15}\,kg} \lesssim m_d \lesssim \mathrm{10^{-14}\,kg}$.  Details of the Helios
measurements are summarized in Table~\ref{tab:data}. 

In addition to 
measuring particle masses and fluxes, the 
instruments  performed a low-resolution compositional analysis with a time-of-flight analyser \citep{auer2001}. 
The Helios dust analyzers were the first 
instruments  measuring  the elemental composition of dust particles in interplanetary space. 

\subsection{Galileo}

Galileo was launched on 18 October 1989, and after two flybys at Earth and one at Venus the spacecraft had enough energy to 
reach Jupiter in December 1995. Galileo was the first spacecraft in orbit about Jupiter until the mission was 
terminated on 21 September 2003. Galileo was a dual-spinning spacecraft, with the dust detector 
 mounted on the despun section of the spacecraft. 
The Galileo dust detector measured dust particle flux, impact direction, speed and mass of the impacting particles 
\citep{gruen1992a}. It was a twin of the dust detector on board Ulysses  \citep{gruen1992b}.

During Galileo's interplanetary mission three orbit segments had a detection geometry which allowed the 
identification of interstellar dust (Figure~\ref{fig:traj_1}). Due to the varying impact speeds, different 
 charge detection thresholds apply to these intervals (corresponding to impact charge 
thresholds ranging from $1\cdot 10^{-13}\,\mathrm{C}$ to $2\cdot 10^{-12}\,\mathrm{C}$,{ \bf cf. Table~\ref{tab:data}}).
 A total of 115 interstellar impactors were 
identified in the Galileo data set \citep{altobelli2005a}.  For our analysis
we have split the three orbit segments shown in Figure~\ref{fig:traj_1} into five time intervals.  
The mass calibration was obtained from an empirical calibration curve 
\citep[][their Figure~3a]{gruen1995a}.

\subsection{Cassini}

\label{sec:cassini}

The Cassini spacecraft was launched on 15 October 1997.  During its first two years in interplanetary space 
the spacecraft performed two flybys at Venus and one at Earth to gain enough energy to reach Saturn. In 2004 it 
became the first spacecraft in orbit about the giant ring planet, until the mission was terminated on 15 September 2017. 
Cassini was a 3-axis stabilized spacecraft.

The Cassini Cosmic Dust Analyzer (CDA)  was an upgrade of the dust detectors flown on board  
Galileo and Ulysses, measuring  particle composition and electric charge in addition to particle mass, 
impact speed, flux and direction \citep{srama2004}. For CDA
we use the mass calibration derived  by \citet[][his Figure~5.1]{stuebig2002}.

Due to operational constraints of the Cassini spacecraft during its interplanetary voyage, 
interstellar dust particles could only be  measured during approximately three months 
from 22~March 1999 to 30 June 1999 at a heliocentric distance between 0.7~AU and 1.2~AU  (Figure~\ref{fig:traj_1}). 
A  charge detection threshold of $3\cdot 10^{-12}\,\mathrm{C}$ had to be used 
in this period, and  $14\pm 3$ particle  impacts of 
likely interstellar origin were identified  in the mass range  
$\mathrm{5\cdot 10^{-17}\,kg} \leq m_d \leq \mathrm{10^{-15}\,kg}$ \citep{altobelli2003}. 
 Details of the Cassini measurements  can be found in Table~\ref{tab:data}. 

\begin{figure}[tb]
\vspace{-10.8cm}
		\includegraphics[width=\textwidth]{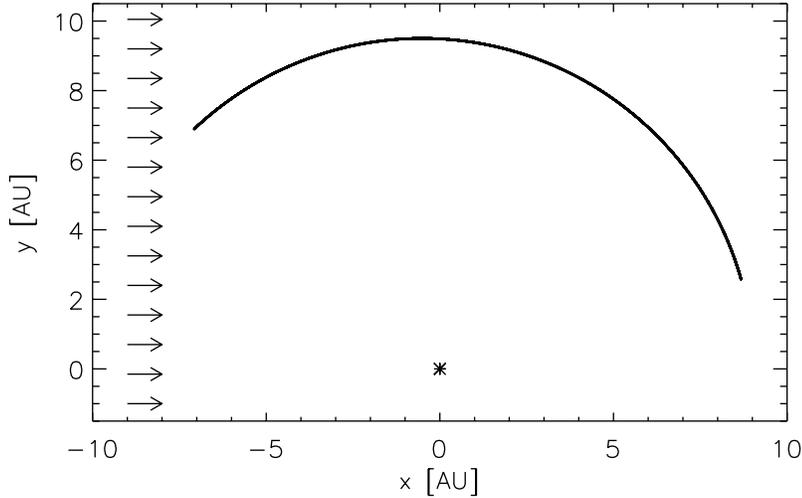}
		\vspace{-2.3cm}
	\caption{Same as Figure~\ref{fig:traj_1} but for Cassini at Saturn. 
	  }
	\label{fig:traj_3}
\end{figure}

The Cassini CDA instrument detected interstellar dust also in the Saturnian system 
\citep[][Figure~\ref{fig:traj_3}]{altobelli2016}.
A total of 36 interstellar particles were identified within a distance range of 9 to 60 Saturn radii from the planet
 by their high entry speed into the Saturnian system  and their impact direction which was compatible with 
the expected interstellar dust flow direction at Saturn. The derived average particle flux was 
$\mathrm{1.5 \cdot 10^{-4}\, m^{-2}\,s^{-1}}$ in the mass range 
$\mathrm{5\cdot 10^{-18}\,kg} \leq m_d \leq \mathrm{5 \cdot 10^{-16}\,kg}$. 

Impacts onto the CDA Chemical Analyzer Target by sufficiently large particles do not provide 
time-of-flight spectra with sufficiently well resolved spectral 
lines from which the minimum impact speed can be derived. The Cassini~2 measurement interval
therefore lacks particles heavier than approximately ${5 \cdot 10^{-16}\,\mathrm{kg}}$. 

 From the rise time of the impact charge signals the particle impact speed can usually be determined with a 
factor of 2 uncertainty. The shape of the time-of-flight mass spectra produced by the CDA Chemical Analyzer Target (CAT),
 however, 
 provides a more accurate determination of the minimum impact speed of each impactor.
 This method, therefore, being based on the 
velocity-mass calibration for the CAT, also provides a better
estimate of the upper mass value for each impactor (compared to the factor of ten, see Section~\ref{sec:data} above).
The time-of-flight mass spectra method also yields a lower dust detection statistics,
compared to measurements performed by the dust instruments on-board Ulysses and Galileo, because of the smaller CAT surface.

\begin{figure}[tb]
\vspace{-11.9cm}
		\includegraphics[width=\textwidth]{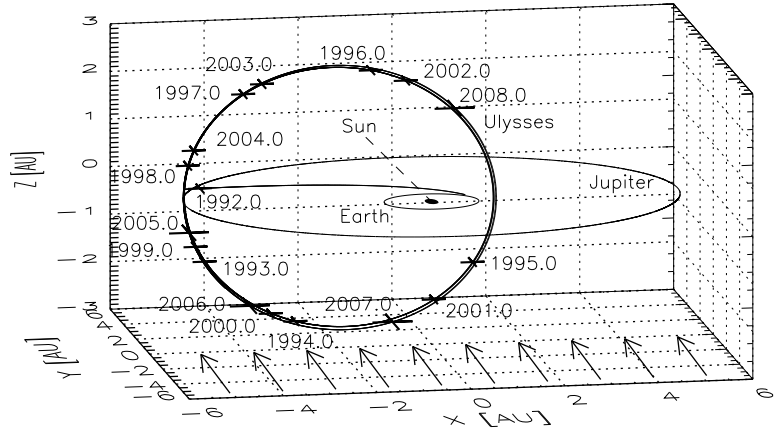}
		\vspace{-1.8cm}
	\caption{Trajectory of Ulysses in ecliptic coordinates with the Sun at the center. The orbits of Earth and Jupiter 
	indicate the ecliptic plane, and the initial trajectory of Ulysses was in this plane. After Jupiter flyby in early 1992, 
	the orbit was almost perpendicular to the ecliptic plane ($79^{\circ}$ inclination). Crosses mark the spacecraft 
	position at the beginning of each year. Vernal equinox is to the right (positive x-axis). Arrows indicate the 
	nominal interstellar dust flow direction, which is within the measurement accuracy co-aligned with the direction of 
	the interstellar helium gas flow. It is almost perpendicular to the orbital plane of Ulysses. 
	  }
	\label{fig:traj_2}
\end{figure}

\subsection{Ulysses} 

Ulysses was launched on 6 October 1990. During  a flyby at Jupiter on 8 February 1992 it was deflected 
on to an orbit almost perpendicular to the ecliptic plane and it became the first spacecraft on a polar orbit about the Sun. 
Operations of Ulysses were terminated on 29 June 2009. The dust detector on board \citep{gruen1992b} 
was a twin of the Galileo dust 
detector. Ulysses was a spinning spacecraft with a period of 5 revolutions per minute.  
The Ulysses trajectory is shown in Figure~\ref{fig:traj_2}.

The Ulysses mission was particularly well suited for the detection of interstellar particles. First, 
its highly inclined orbit  with an aphelion at approximately Jupiter's orbit (5.5~AU) took the spacecraft far 
above the ecliptic plane. 
Given that the concentration of interplanetary dust particles drops at increasing ecliptic latitudes and 
that most of the interplanetary dust moves on prograde heliocentric orbits, the orbital sections where  
Ulysses was far from its perihelion and far from  the solar poles were best suited for the detection of interstellar dust. 
The orientation of Ulysses's orbital ellipse was such that in these sections the impact 
directions of interplanetary and  interstellar particles were almost antiparallel, and thus these populations could
 easily be separated. 
 
 The Ulysses  dust data  is by far the largest data set 
of in-situ interstellar dust measurements available to date. A detection threshold of 
$1\cdot 10^{-13}\,\mathrm{C}$ had to be used and the data set contains more than 900 identified interstellar 
particles, covering about 75\% of one full 22-year solar cycle \citep{strub2015,krueger2010b,krueger2015a}. 
For our analysis we have 
selected five mission intervals when the dust detector was continuously measuring dust (cf. Table~\ref{tab:data}). 
. Similar to Galileo, the mass 
calibration was obtained from an empirical calibration curve \citep[][their Figure~3a]{gruen1995a}.
Given the long time coverage and thus large number of identified interstellar particles,
the  dust model we are going to use in 
Section~\ref{sec:model} was  calibrated with the Ulysses interstellar dust data set \citep{strub2019}.

\section{Interstellar Dust Simulations}

\label{sec:model}

Previous models for interstellar dust in the solar system described  the interstellar dust flow
at larger heliocentric distances well, but they did not have the resolution to enable a good time-resolved 
description of the dust environment at Earth \citep{gruen1994a,landgraf2000b,sterken2012a}. 
Based on these earlier models and the dust measurements by the Ulysses spacecraft, \citet{strub2019} 
executed high-resolution simulations in the context of the IMEX modelling effort (Interplanetary Meteoroid 
environment for EXploration) that included an interstellar dust module developed for this purpose. 
The authors simulated the dynamics of charged micrometer and sub-micrometer sized interstellar particles 
exposed to solar gravity, 
solar radiation pressure and a time-varying IMF. 
The mass distribution is 
represented by 12 particle sizes 
between $\mathrm{0.049\,\mu m}$ and $\mathrm{4.9\,\mu m}$, and the dynamics of each of these sizes was 
simulated individually. 

In IMEX the dust density in the solar system
is calibrated with the Ulysses interstellar dust measurements which is by far the most comprehensive 
data set of interstellar dust measurements presently available \citep{strub2015,krueger2015a}. Each particle size
bin in the model was calibrated such that the average dust flux measured by
Ulysses  in this size bin was reproduced \citep{strub2019}. 
Due to the variation of the IMF imposed by the 22-year solar cycle, the  model
is time-dependent. For details of the  model  and general interstellar dust flow characteristics the reader is 
referred to \citet{sterken2012a}, and  for  the flow at Earth orbit to \citet{strub2019}.
Here we use  IMEX  to simulate dust fluxes and we compare the results with the dust measurements 
discussed in Section~\ref{sec:data}. 

 \begin{figure*}[htb]
\vspace{-4.5cm}
	\hspace{-0.8cm}
		\includegraphics[width=1.25\textwidth]{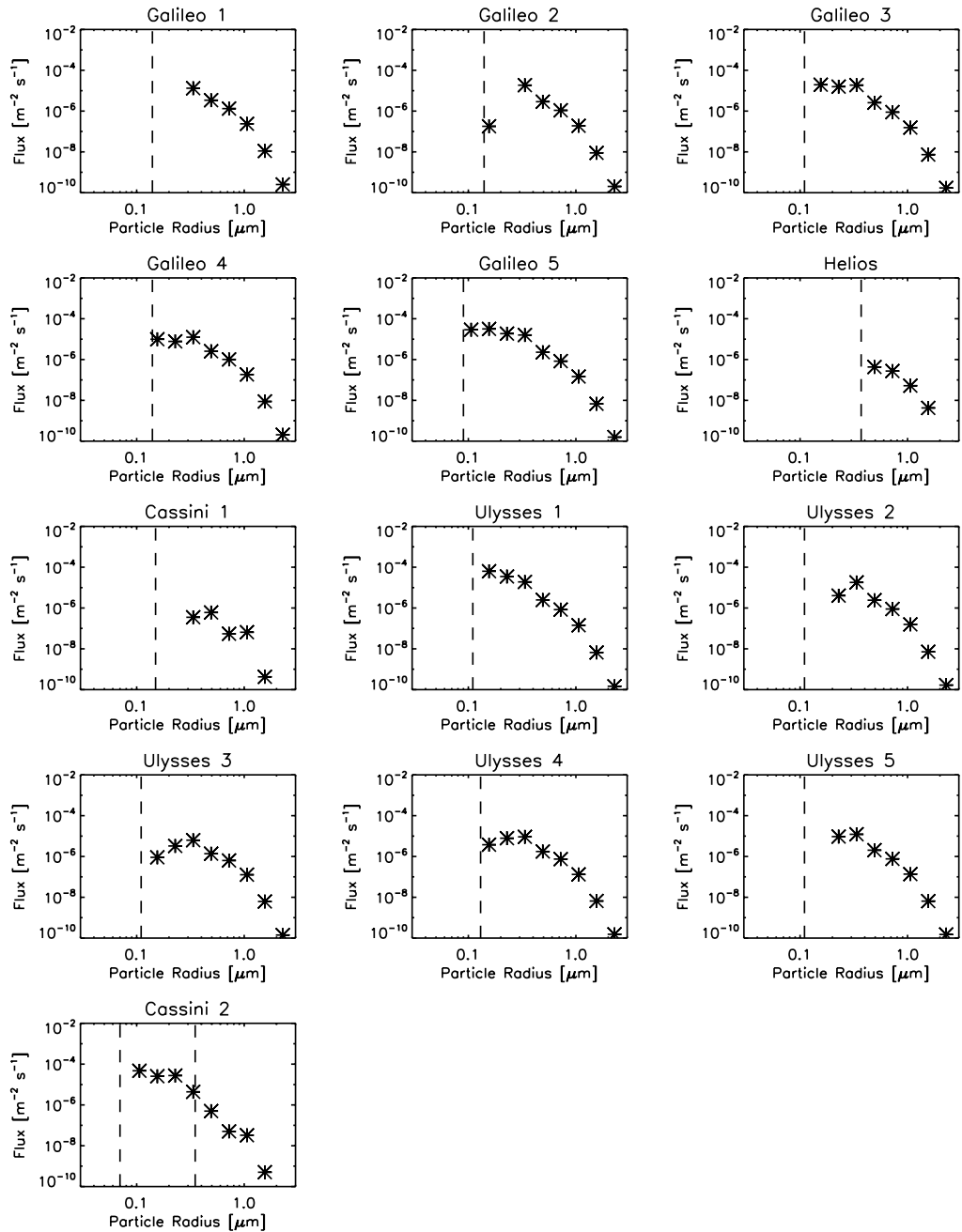}
		\vspace{-3.1cm}
	\caption{Interstellar particle size distributions obtained from the simulations for the different
	missions and orbit segments given in Table~\ref{tab:data}. Dashed vertical lines indicate the  detection thresholds of the 
	dust instruments assuming an average particle impact speed derived from the model, as discussed 
	in Section~\ref{sec:data} (columns~5 and 9 in Table~\ref{tab:data}). For the Cassini~2 interval the upper 
	detection limit  at approximately $0.35\,\mu \mathrm{m}$ particle radius for the CDA Chemical Analyzer Target is also shown. 
	  }
	\label{fig:flux_1}
\end{figure*}

\begin{figure*}[htb]
	\vspace{-3.cm}
	\hspace{-0.1cm}
		\includegraphics[width=1.1\textwidth]{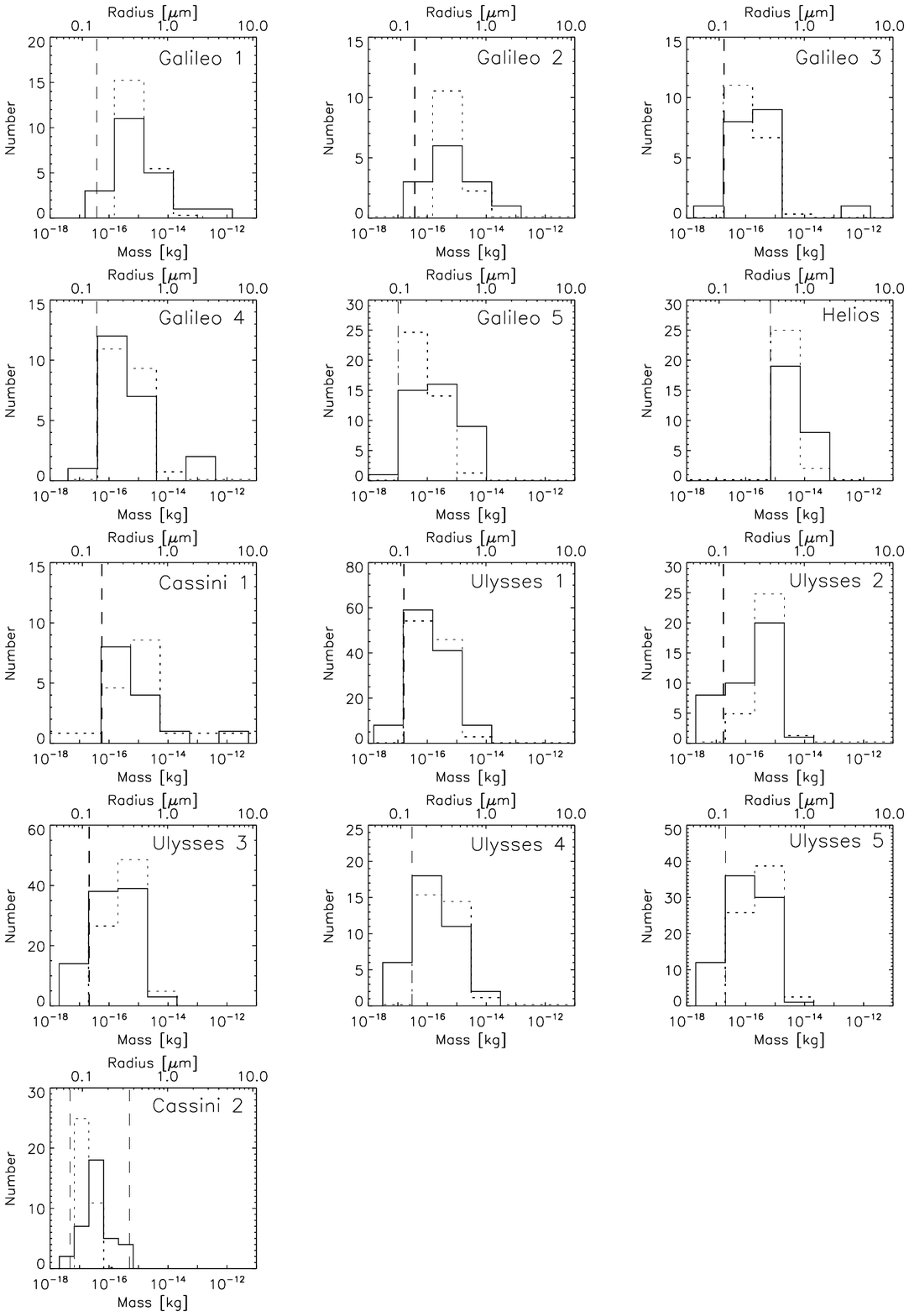}
		\vspace{-3.2cm}
	\caption{Mass distributions of interstellar impactors derived from the instrument calibration with the 
	calibration parameters 
	listed in Table~\ref{tab:data} (solid histograms; from \citet{altobelli2005a,altobelli2003,altobelli2006,altobelli2016,strub2015}). 
   Dashed histograms show the fluxes shown in Figure~\ref{fig:flux_1} derived from the model.  The model curves 
	are normalized such that they contain the same number of particles as measured by the spacecraft detector in the time
	interval under consideration. 	Vertical dashed lines indicate the  detection threshold 
	assuming an average particle impact speed derived from the model (columns~5 and 9 in Table~\ref{tab:data}); 
	for the Cassini~2 interval the upper detection limit for the CDA Chemical Analyzer Target is also shown. 
	Particle radii are indicated at the top. 
	  }
	\label{fig:massdist}
\end{figure*}

Throughout this paper we use the inflow direction of the  
interstellar dust into the heliosphere from a direction of $259^{\circ}$ ecliptic longitude and 
$8^{\circ}$ latitude  and  inflow speed 
of $\mathrm{26\,km\,s^{-1}}$  as the nominal direction and
speed of the  undisturbed interstellar dust flow.
This is equivalent to the interstellar particles being at rest with respect to the local interstellar
cloud, and they approach the Sun on hyperbolic trajectories. The trajectories of 
particles with $\beta=1$ are  altered neither by solar radiation nor by solar gravity. The 
only force leading to a deflection of these particles is the Lorentz force imposed by the IMF. 
For particles with $\beta \not= 1$ the solar radiation pressure 
leads to either a concentration of particles ($\beta  < 1$) downstream of the Sun, or to a deflection 
($\beta > 1$) and thus to a time-independent depletion cone downstream of the Sun (Figure~\ref{fig:traj_1}). 

In our model,
the dynamics of particles with approximate radii $\mathrm{0.1\,\mu m} \lesssim r_d \lesssim \mathrm{0.5\,\mu m}$ 
are dominated by solar radiation pressure, 
larger particles
are dominated by gravity, and smaller particles by the Lorentz force imposed by the IMF. 
The $\beta$-mass relation used for each particle size in our simulations is determined according to the 
``adapted astronomical silicates $\beta$-curve''~\citep{sterken2012a} which combines the radiation pressure 
efficiencies of \citet{gustafson1994} with an average of the maximum values for $\beta$ from Ulysses observations 
\citep[$\beta = 1.6$, from ][]{landgraf1999a}.

The time-variable IMF evolves through focussing and defocussing configurations 
during the 22-year solar cycle. This causes time-dependent concentrations and rarefactions of 
interstellar particles smaller than approximately $\mathrm{0.3\,\mu m}$ in the inner solar system.
We use the solar cycle approximation used by \citet[][their Figure~1]{strub2019}. Around 
the solar minima in 1974 and 1996, the IMF was in a defocussing configuration and around the  solar minima in 
1985 and 2007 the IMF was in a focussing configuration for these small particles. 

We did not take into account the  sensitivity profiles of the individual dust instruments. Instead, the
model assumes a spherical sensor with $4\pi$ sensitivity characteristics, i.e. all particles reaching the
spacecraft are  taken into account. Given that for all space missions under consideration the
dust sensors had a rather wide field-of-view and the interstellar dust flow is rather collimated,
this is a reasonable approximation.

\subsection{Dust Size Distributions}

\label{sec:size_dist}

The simulated interstellar dust size distributions for the  measurement periods of the four spacecraft 
considered here are shown in Figure~\ref{fig:flux_1}. 
Vertical dashed lines indicate the  detection thresholds which had 
to be applied   
to separate interstellar particles from other dust populations in the data sets 
\citep{altobelli2003,altobelli2005a,altobelli2006,altobelli2016,strub2015}. Thus, interstellar particles
to the left of the dashed lines -- if present -- could not be extracted from the 
data. Therefore, to avoid confusion in the following discussion, we do not show particles  
in this size range in Figure~\ref{fig:flux_1} even though they may be present in the model. 

Strong variations  are  
imposed by the varying heliocentric 
distances of the spacecraft and the time-dependent IMF configuration. The smallest particles are most  effectively 
prevented from entering the inner solar system  during defocussing configurations of the IMF. This leads
to a strong depletion of  particles smaller than approximately $\mathrm{0.3\,\mu m}$,
in particular when the Helios and Cassini measurements were taken in 
the innermost regions of the solar system (cf. Table~\ref{tab:data}). 
Furthermore, particles in the size range $\mathrm{0.1\,\mu m} \lesssim r_d \mathrm{\lesssim 0.5\,\mu m}$
are depleted by the solar radiation pressure. 
Similarly, the Ulysses measurements
2 to 5 were taken during the 
defocussing configuration  
and are thus also strongly depleted in small particles. On the
other hand, the Galileo intervals 1 to 3 were in the focussing phase of the IMF when 
small particles could reach the inner solar system.

In Figure~\ref{fig:massdist} we compare the simulated dust mass distributions with the spacecraft 
measurements (the conversion between particle mass and radius is done with Equation~\ref{equ_2}, see 
footnote of Table~\ref{tab:data}). Again, the  detection thresholds are indicated (Table~\ref{tab:data}, column~8). 
Particle masses were calculated from the measured impact charges with the 
average particle impact speeds derived from the simulations (Table~\ref{tab:data}, column~5), 
in the same way as the instrument detection thresholds.

\begin{figure}[tb]
	\vspace{-5.5cm}
		\includegraphics[width=\textwidth]{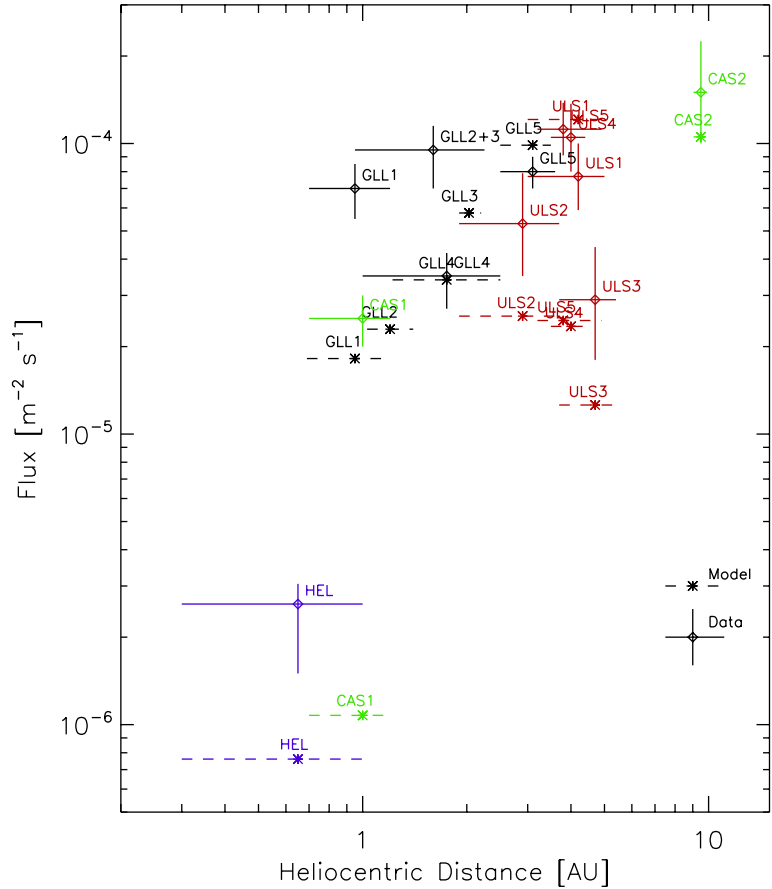}
		\vspace{-2.5cm}
	\caption{Dust fluxes simulated with IMEX (asterisks) and measured (diamonds) with Helios (HEL), Galileo (GLL),
	Cassini (CAS) and Ulysses (ULS). Horizontal bars indicate the distance range where dust measurements and simulations were 
	performed. The detection thresholds and measured particle size ranges listed in Table~\ref{tab:data} and 
	shown in Figure~\ref{fig:flux_1} have been adopted for the simulations.  
	  }
	\label{fig:flux_2}
\end{figure}

In general, the  masses predicted by the IMEX model are in good agreement with the
measurements, only the Ulysses data show some impacts below the calculated detection thresholds. 
This is likely due to uncertainties in the mass calibration and/or the detection threshold. Note that, 
 particle
masses were derived from the measured impact charge using Equation~\ref{eq_charge} with a typical factor of 10
uncertainty in the mass calibration.
Furthermore, 
all impacts were calibrated with the same average impact speed derived from the model, while the model
predicts a speed variation of about a factor of two during each of the measurement intervals. For the
second Cassini interval, the minimum impact speed was derived from the impact spectra with a typical 
accuracy of 5\,\kms, leading to a higher accuracy
in the mass calibration. Thus, the data in Figure~\ref{fig:massdist} are binned in intervals with one
order of magnitude bin width, except for Cassini~2 where we achieved a higher mass resolution 
(cf. Section~\ref{sec:cassini}).

\subsection{Dust Fluxes}

\label{sec:fluxes}

\begin{figure}[tb]
	\vspace{-7.8cm}
		\includegraphics[width=\textwidth]{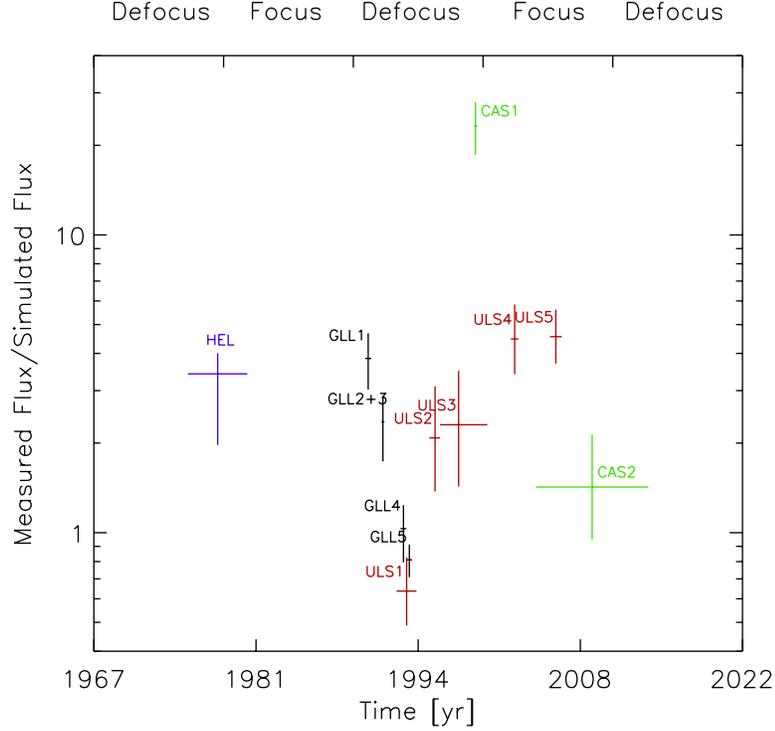}
		\vspace{-2.5cm}
	\caption{Ratio of  measured to simulated dust fluxes from Figure~\ref{fig:flux_2}. 
	 Horizontal bars indicate the time intervals when dust measurements and simulations were 
	performed. The modelled focussing and defocussing phases of the IMF in the inner solar 
	system based 
	on observations by the Wilcox Solar Observatory \citep{hoeksema2018} are indicated at the top 
	\citep{strub2019}. 
	  }
	\label{fig:flux_ratio}
\end{figure}

 In Figures~\ref{fig:flux_2} and \ref{fig:flux_ratio} we show the simulated dust fluxes 
integrated over all detectable particle sizes and compare them with the measurements.  We added up  
simulated fluxes for  sizes only above the detection threshold shown in 
Figure~\ref{fig:flux_1} because smaller particles were not detectable by the dust instruments. 
The simulations
were performed for the same time intervals as the measurements (columns~2 and 3 in Table~\ref{tab:data}).
All measured and simulated dust fluxes discussed in this paper are given in the heliocentric 
reference system.

The model predicts on average somewhat lower fluxes 
than measured.  This is also the case for Ulysses, even though the Ulysses data were 
used to calibrate the IMEX 
model.  We will come back to this in Section~\ref{sec:discussion}. The mean value of the 
ratio between measured and  
simulated fluxes for Galileo, Cassini and 
Helios is  $\mathrm{2.6^{+5.5}_{-1.7}}$ (for all measurements including Ulysses 
it is  $\mathrm{2.5^{+4.0}_{-1.5}}$, $1\sigma$ uncertainty).  If we ignore the
data point for Cassini~1 we get a mean value of
 $\mathrm{1.8^{+1.7}_{-0.8}}$ (without  Ulysses).
 
\section{Discussion}

\label{sec:discussion}

The IMEX model has been calibrated with the Ulysses interstellar dust data set because it is the most 
comprehensive and homogeneous measurement by a single dust instrument over a period of 16 years, covering a large 
portion of a full 22-year solar cycle (Figure~\ref{fig:flux_ratio}). With more than 900 
identified interstellar particles it has the best statistical accuracy of all interstellar dust measurements
with a dedicated dust detector obtained to date. For three of the five  
Ulysses measurement intervals the model reproduces 
the data within  a factor of 2. Only after 2002 is the discrepancy  somewhat more pronounced.

Given that the Ulysses data set was used to calibrate the IMEX model, this discrepancy is surprising 
at first glance. However, the model was calibrated with the full Ulysses data set, and it reproduces the
overall interstellar dust fluxes measured during the entire mission to within 2\%. 
This makes us very confident that the overall calibration of the model consistently reproduces the Ulysses measurements. 

For the analysis in this paper
we have used five relatively short time intervals for Ulysses instead of the full mission data set, mainly for 
two reasons: (1) we wanted to have a comparable number of dust impact events in each time interval as is the
case for
the other missions considered in this paper, and (2)  we wanted to cover shorter heliocentric distance 
ranges than the entire Ulysses mission. Furthermore, we ignored the spatial region in the 
vicinity of Jupiter to make sure that we do not have a contamination by Jupiter stream particles \citep{krueger2006c}.
Thus, we disregarded the time period from 2003 to mid 2005. 
It is not surprising that we get a somewhat larger discrepancy between model and
data for shorter time intervals even though the model agrees very well with the data for the full time period. 

The IMEX model reproduces most of the Galileo, Helios and Cassini measurements to within a factor
of 2 to 3. Only during the Cassini~1 interval is the measured flux  larger by approximately a factor of 20 
 then predicted by the model. It should be noted that the model as calibrated with the Ulysses data 
 shows a tendency to predict  {\em lower}
fluxes than measured by the space instruments.  

The underestimation of the  flux in the Cassini~1 time interval  may be related to the representation of the IMF by a Parker model 
\citep{strub2019}: the Parker model describes the quiet IMF during solar minimum rather well, 
the Cassini~1 measurements, however,  were performed around solar maximum during the strongest IMF defocussing conditions. 
Large deviations from the  Parker IMF have to be expected, for example, inside CMEs which are not accounted for by the
model. They can lead to severe discrepancies with the simulated flux. 
 
Furthermore, discrepancies between model and data may be related to the demonstration by \citet{sterken2015} that either 
the first part of the data (before 2002) or the last part (2002-2008) 
can be well represented by simulations of one dust population (size and physical properties) but not both periods in one simulation
with a single set of  particle properties. In 2005 a rapid change in interstellar flow direction and density was seen in the 
Ulysses data \citep{krueger2007b}, while in 2006 the
flow direction was again co-aligned with the nominal flow direction
of the interstellar helium within the measurement accuracy \citep{strub2015}. The authors concluded that this was 
a temporally limited 
phenomenon. \citet{sterken2015} simulated this shift in dust direction (i.e. the data after 2002) for porous bigger 
particles ($\mathrm{\gtrsim 0.2\,\mu m}$ radius) and for rather compact smaller ($\mathrm{\lesssim 0.2\,\mu m}$) particles. 
An extra (solar-cycle dependent) filtering mechanism 
in the outer boundary regions of the heliosphere \citep{kimura1998} included in the simulations may lead to a better fit for the entire Ulysses data set,
in particular to reproduce the shift in dust flow direction in 2005.  

The Ulysses measurements which were used for the model calibration covered the time period from 1992
to 2007, a spatial region between approximately 2~AU and 5~AU, and they were acquired most of the time far 
away from the ecliptic plane  \citep{strub2015,krueger2015a}. The measurements by the other spacecraft 
were obtained in the ecliptic plane, at  distance ranges much 
closer to the Sun inside Earth's orbit (Helios, Galileo and Cassini~1),  further away from the Sun 
 at Saturn (Cassini~2), or many years before Ulysses (Helios). 
The model shows  overall good agreement with all these dust measurements, in particular those  
obtained in spatial regions not used for the model calibration.

All interstellar dust measurements obtained by Helios, Galileo and Cassini were obtained in the ecliptic plane 
in environments 
where solar-system dust populations dominate the particle fluxes: interplanetary
dust particles dominate during all  measurement intervals of these missions, 
while Saturnian dust makes an additional significant contribution in the Cassini~2 interval. 
Only Ulysses had an ideal configuration for 
interstellar dust detection far away from the ecliptic plane most of the time. Therefore, the
interstellar dust measurement of Helios, Galileo and Cassini are connected with larger uncertainties in the
particle identification and, hence, dust fluxes than  those of Ulysses.  

For Ulysses the selection criteria for the identification of interstellar particles in the data set used in this work
are the same as the ones that were used by \citet{strub2015} to analyze the dynamical properties of the particles. 
\citet{krueger2015a} used different criteria to derive the mass distribution of the particles 
in order  not to induce any bias in the mass distribution. With their technique, these authors identified 
interstellar particles as small as $\mathrm{2\cdot 10^{-18}\,kg}$ in the Ulysses data set. 
It indicates that even though such small particles are strongly filtered by the heliospheric interaction, a 
fraction of them can still reach the inner solar system between 2~AU and 5~AU. 
 
In Figure~\ref{fig:massdist} the mass distributions measured by Ulysses are well reproduced by the model. 
On the other hand, the model overestimates the abundance of small particles close to the detection thresholds for a few 
of the Galileo measurement intervals and for Helios. In the future we may 
include these other dust measurements (Galileo, Cassini and Helios) to calibrate the model. This may 
improve the overall agreement between model and data for the dust fluxes and mass distributions. It is, 
however, beyond the scope of our present paper. 

The Cassini measurements at Saturn (Cassini~2) show a deficit of  small particles with masses 
below approximately $2\mathrm{\cdot 10^{-17}\,kg}$ (Figure~\ref{fig:massdist}), despite the fact that 
the sensitivity of the instrument
enables the detection of smaller particles down to  $\mathrm{5\cdot 10^{-18}\,kg}$ \citep{altobelli2016}. 
This can be explained by the filtering of such small particles at the heliopause and the inner heliosphere 
\citep{sterken2013a,slavin2012}, a phenomenon  also observed in the Ulysses interstellar dust data 
\citep{landgraf2000b}. Furthermore, 
\citet{altobelli2016} confirm the existence of particles with $\beta > 1$ with a maximum value reached between 
$\mathrm{10^{-17}\,kg}$ and $\mathrm{10^{-16}\,kg}$, in good agreement with the $\beta$-mass domains 
inferred from the Ulysses data \citep{landgraf1999a,kimura2003b}. 
 
The lack of large interstellar particles  in the Cassini~2 data is due to the detection method
on the CDA Chemical Analyzer Target (CAT): large impacts, typically
micron-sized particles, do not provide time-of-flight spectra with sufficiently well resolved spectral 
lines from which the
minimum impact speed can be derived. Furthermore, the almost ten times smaller CAT target area compared to the
detection area of the Ulysses and Galileo instruments  strongly reduces the likelihood of large particle detections.

Figure~\ref{fig:flux_2} shows an overall increase in the dust fluxes as a function of heliocentric distance. 
The detection thresholds varied for the different missions and measurement 
intervals, nevertheless this trend illustrates the filtering of the interstellar dust particles by the 
heliosphere. It confirms the 
earlier results by \citet{altobelli2005b} which were based on a smaller data set and on the measurements alone, i.e.
without modelling. 
 
Finally, the ratio between measured and simulated fluxes in Figure~\ref{fig:flux_ratio} 
does not show a systematic trend with the solar cycle. It indicates that the description of the 
heliospheric filtering by the IMF implemented the model is rather reliable.  

The decrease in measured vs. simulated flux around 1994 may be caused by the filtering effect of the heliospheric 
boundary which is not yet implemented in the model. This was illustrated by \citet[][their Fig.~19]{sterken2015}: 
Particles passing through the solar system in 1994 have passed the boundary regions of the heliosphere in the 
defocusing phase of the solar cycle. In this region, higher particle charges \citep{kimura1998,slavin2012} 
lead to larger Lorentz forces,  thus filtering out interstellar dust pacticles,  despite of a lower magnetic 
field strength in comparison with the solar system IMF. 
While \citet{sterken2015} suggested this hypothesis for explaining the Ulysses data, here also the 
Galileo data (GLL4 and GLL5) seem to follow this trend. Further analysis is needed for confirmation.  

The overall agreement between model and data indicates that an extrapolation of the model in space and time should, in 
general,  give reliable predictions for future space missions  with a tendency to underestimate the 
expected dust fluxes. 
The IMEX model was recently used to study the dust detection conditions for the  DESTINY$^+$ mission
which will measure dust in interplanetary space between 2024 and 2028 and during a dedicated 
flyby at the active asteroid (3200) Phaethon \citep{kawakatsu2013,krueger2019a,kimura2019,szalay2019}.

\section{Summary}

\label{sec:conclusions}

We have used the interstellar dust module of the Interplanetary Meteoroid environment for EXploration model 
\citep[IMEX;][]{sterken2013a,strub2019} to simulate the dynamics of interstellar dust in the solar system. The model covers all relevant 
forces, i.e. solar gravity, solar radiation pressure, and electromagnetic interaction with the 
interplanetary magnetic field. We have compared our model results with in-situ interstellar 
dust measurements obtained with four spacecraft, i.e. Helios, Galileo, Cassini, and Ulysses 
\citep{altobelli2006,altobelli2005a,altobelli2003,altobelli2016,strub2015}. Our results
can be summarized as follows: 

The model gives  overall good agreement with the spacecraft measurements. 
Dust fluxes and size distributions simulated for  time intervals and spatial regions not covered in the
original calibration of the model agree with the in-situ spacecraft measurements to within a factor of 2 to 3. 
This marks the limit of our current understanding of the interstellar dust flow through the solar system.
The model usually underestimates the dust fluxes measured by spacecraft. 
 
IMEX is a unique time-dependent model for the prediction of interstellar dust fluxes and
mass distributions for the inner and outer solar system. The model is suited to study 
dust detection conditions for past and future space missions.

\section*{Acknowledgements}
 
The IMEX model was developed unter ESA funding (contract 4000106316/12/NL/AF - IMEX). H.K. and P.S. are grateful to the 
MPI f\"ur Sonnensystemforschung and the University of Stuttgart for their support.  We are grateful to an
anonymous referee  whose comments substantially 
improved the presentation of our results.


\end{document}